\documentclass[fleqn,twocolumn,10pt]{wlscirep}

\usepackage[utf8]{inputenc}
\usepackage[T1]{fontenc}

\usepackage[normalem]{ulem}
\usepackage[colorlinks=true,allcolors=blue]{hyperref}
\usepackage{amsmath}
\usepackage{mathtools}
\usepackage{upgreek}
\usepackage{braket}
\usepackage{dcolumn}
\usepackage{xcolor}
\usepackage{graphicx}
\usepackage{orcidlink}
\usepackage{xfrac}
\usepackage[section]{placeins}

\newcommand{\Baq}[1]{\ensuremath{^{#1}\text{Ba}^+}}

\newcommand{\Tm}[1]{\ensuremath{^{#1}\text{Tm}}}
\newcommand{\Tmq}[1]{\ensuremath{^{#1}\text{Tm}^+}}

\title{Hyperfine spectroscopy of optical-cycling transitions in singly ionized thulium}

\author[1]{Patrick M\"{u}ller\orcidlink{0000-0002-4050-1366}}
\author[1]{Andrei Tretiakov\orcidlink{0000-0001-9615-8563}}
\author[1,a]{Amanda Younes\orcidlink{0009-0004-1985-6113}}
\author[1,b]{Nicole Halawani\orcidlink{0009-0005-7948-9658}}
\author[1,*]{Wesley C. Campbell\orcidlink{0000-0003-1578-906X}}
\author[1]{Paul Hamilton\orcidlink{0000-0001-6080-632X}}
\affil[1]{Department of Physics and Astronomy, University of California Los Angeles, Los Angeles, CA 90095, USA}

\affil[a]{Present address: Department of Physics, Harvard University, Cambridge, MA 02138, USA}
\affil[b]{Present address: Department of Physics, University of California Santa Barbara, Santa Barbara, CA 93106}

\affil[*]{Corresponding author: Wesley C. Campbell (wesley.campbell@physics.ucla.edu)}

\keywords{Thulium-169 ion spectroscopy, Hyperfine structure, Optical cycling transitions, Laser cooling of ions, Metastable qubit state, Ion trap experiments}

\begin{abstract}
We present a spectroscopic investigation of \Tmq{169} that provides two key foundations for its use as a platform for advanced quantum applications. First, we establish the complete spectroscopic road map for optical cycling (including laser cooling) by performing high-resolution spectroscopy on \Tmq{169} ions in an ion trap. We characterize the primary $313$\,nm and complementary $448/453$\,nm cycling transitions, identify the essential near-infrared repumping frequencies, and determine the magnetic-dipole hyperfine $A$ constants for all relevant levels. Second, we report a detailed characterization of a metastable state as a candidate for hosting a robust qubit, performing lifetime measurements and Zeeman-resolved microwave hyperfine spectroscopy with $\mathrm{kHz}$ precision.
\end{abstract}
\begin{document}

\flushbottom
\maketitle

\thispagestyle{empty}

\section*{Introduction}
\label{sec:introduction}

Despite the maturity of neutral-atom quantum platforms across the periodic table\,\cite{McClelland.2016}, including alkalis\,\cite{Isenhower.2010, Jau.2016}, alkaline-earth-like atoms\,\cite{Madjarov.2020, Ma.2022}, rare-earth metals\,\cite{McClelland.2006, Sukachev.2010, Lu.2011, Dutta.2014}, metastable noble gases\,\cite{Vassen.2012}, and group\,III elements\,\cite{Yu.2022}, laser cooling and control in trapped ions have been mostly focused on alkaline-earth and alkaline-earth-like ions with an $ns\ {}^{2}\mathrm{S}_{1/2}$ ground state\,\cite{Bruzewicz.2019}. While the level structure of such elements offers relatively convenient schemes for laser cooling, state preparation and measurement (SPAM), they lack a large number of accessible internal states and multilevel schemes. In contrast, additional long-lived states in non-alkali-like ions could enable architectures that partition roles across distinct qubits (e.g., processing vs. memory), as proposed in the optical, metastable, and ground-state ($omg$) qubit protocol\,\cite{Allcock.2021, Vizvary.2024}. While the feasibility of laser cooling of open-$4f$-shell rare-earth ions has been discussed before\,\cite{Lepers.2016}, no real experimental efforts for laser cooling have been demonstrated. Here we present such efforts in \Tmq{169}.

Thulium is a rare-earth lanthanide metal with a single stable isotope, $\Tm{169}$. Neutral \Tm{} has recently been laser cooled\,\cite{Sukachev.2014, Provorchenko.2023} and used for atomic clocks\,\cite{Sukachev.2016, Golovizin.2019}, Bose--Einstein condensates\,\cite{Khlebnikov.2019, Davletov.2020, Tsyganok.2023}, and as a quantum-computing platform\,\cite{Mishin.2025}. 
\Tmq{169} in its ground state has a single hole in the $4f$ shell and one electron in the $6s$ shell, generating the $4f^{13}6s\,{}^3\mathrm{F}_4^{\mathrm{o}}$ state, or $(\sfrac{7}{2},\sfrac{1}{2})^{\mathrm{o}}_4$ in the $J_1j$-coupling basis\,\cite{NIST}. This configuration yields manifolds of relatively low-lying states with total electronic angular momentum spanning from $J=1$ to $J=9$. Such large-$J$ manifolds are attractive for absorption--emission code schemes that require at least $J\ge \sfrac{9}{2}$\,\cite{Jain.2024, Aydin.2025}. At the same time, the nuclear spin $I=\sfrac{1}{2}$ ensures that each fine-structure level splits into only two hyperfine-structure (HFS) states, simplifying laser cooling and qubit initialization by reducing the number of frequency tones required for closed population transfer.
\begin{figure*}[!t]
\centering
\includegraphics[width=\textwidth]{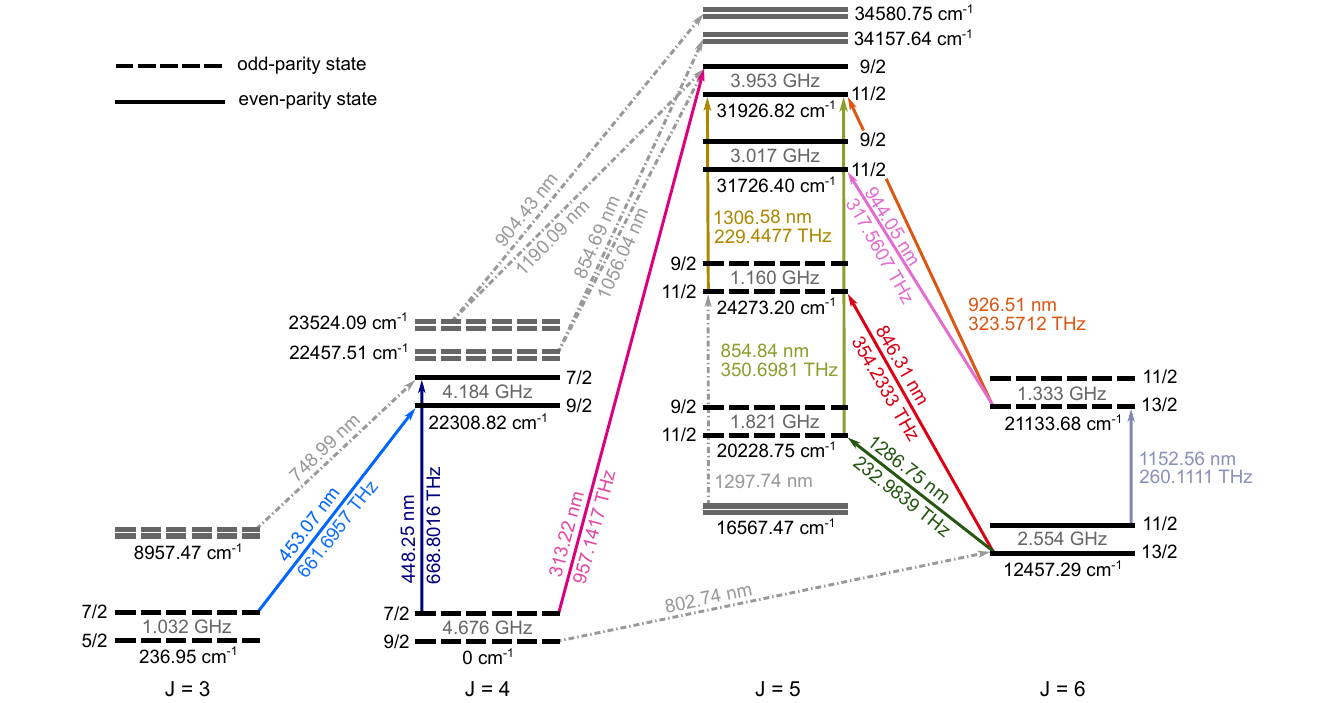}
\caption{\textbf{\textit{E}1 transitions and hyperfine splittings in \Tmq{169}.} Levels with the same $J$ value are aligned vertically, with $F$ values labeling the hyperfine sublevels. Level energies are indicated in cm$^{-1}$ and the measured hyperfine splittings are given in GHz. The level spacings are not to scale, and levels irrelevant to this work are omitted. The vacuum wavelength and corresponding frequency are shown next to one of the two strongest hyperfine components ($\Delta F = \Delta J$) of each transition. Gray levels and gray dash-dotted lines indicate levels and expected transitions, respectively, that were searched for but not observed in this work.}
\label{fig:levels}
\end{figure*}

For this study, two optical cycling schemes were investigated that have cooling transitions at wavelengths of $450$\,nm and $313$\,nm. These were selected based on their transition strengths, leakage channels and the accessibility of the required wavelengths with current laser technology. The energy levels and $E$1 transitions investigated and relevant for laser cooling and SPAM are depicted in Figure~\ref{fig:levels}. In this diagram, solid arrows indicate transitions observed experimentally, while dashed arrows denote potential decay channels predicted by theoretical calculations but not observed in our spectroscopic measurements. This level diagram is non-exhaustive; for clarity, we omit some intermediate levels and those with $J < 3$ and $J > 6$. For each wavelength, the measured frequency of one of the two strongest hyperfine components ($\Delta F = \Delta J$) is indicated together with the corresponding inferred level splittings.

Of the spectroscopic data on \Tmq{169} that has been collected experimentally\,\cite{Martin.1978, Mansour.1989, Wang.2022, Kebapci.2024, DenHartog.2024}, only three transitions relevant to our proposed cooling schemes have been reported in the literature: $448.25$\,nm, $453.07$\,nm, and $313.22$\,nm, with no data on the transitions corresponding to possible repumping transitions. Guided by theoretical predictions for possible decay channels\,\cite{Quinet.1999,Quinet.DREAM,Palmeri.2025}, we measured the transitions' wavelengths and HFS of the relevant levels.

We employed three different experimental setups to determine the HFS of the levels involved in the proposed cycling schemes. Our most precise measurements were obtained from laser-induced fluorescence on a trapped ion cloud, while preliminary spectra were recorded using a hollow-cathode lamp (HCL) and a small vacuum chamber with an ablation target. The latter two methods can be implemented quickly and cost-effectively compared with an ion-trap apparatus and are therefore well-suited for evaluating the feasibility of working with a new ion species before committing to modifications of an established trap setup. The experimental methods are described in detail in Methods. The following introduction of the proposed laser-cooling cycles is presuming an ion trap environment.

\subsection*{Laser cooling at 450\,nm}
\label{sec:laser cooling at 450 nm}
\begin{figure}[!t]
\centering
\includegraphics[width=\linewidth]{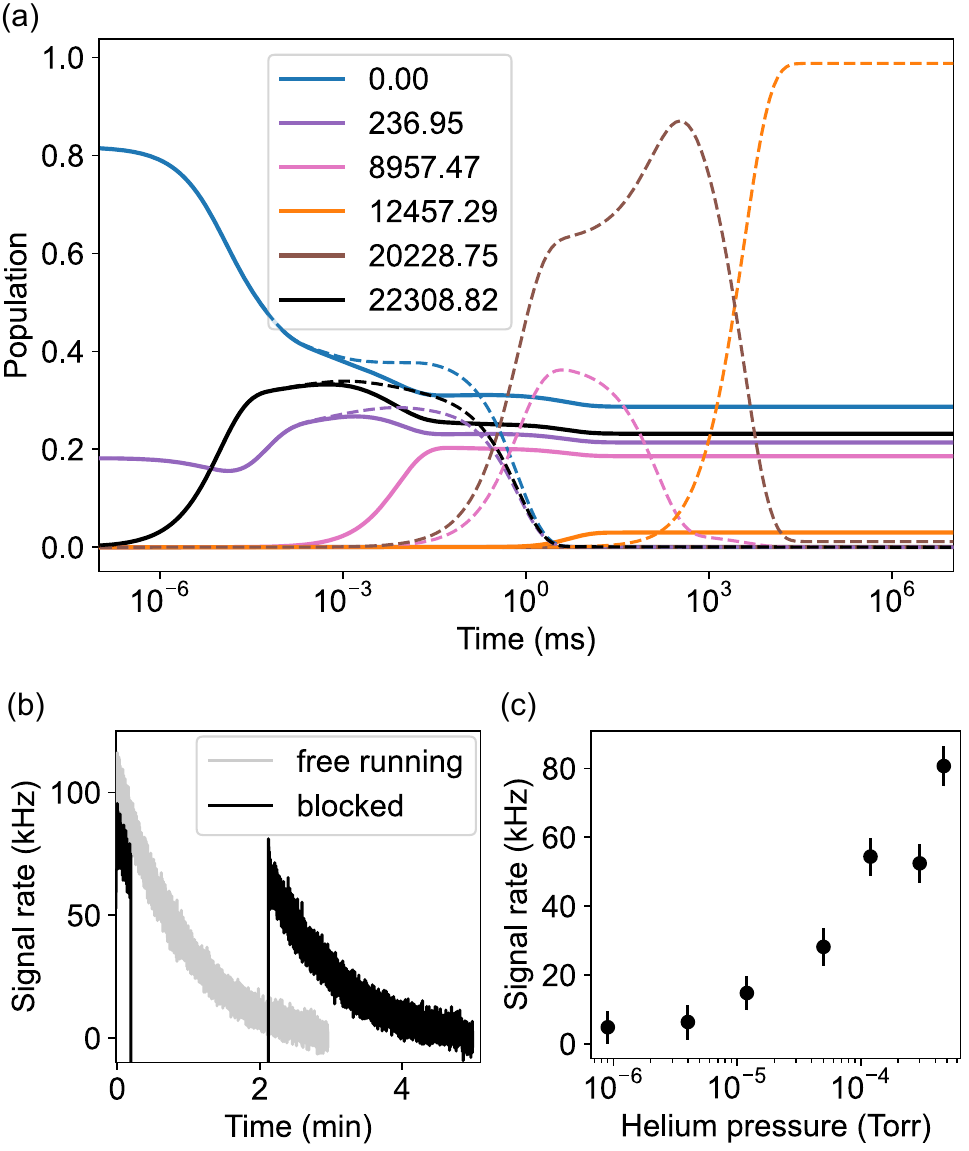}
\caption{\textbf{Population dynamics in \Tmq{169}.} (a) Solution of the rate equations for the 450-nm cycle. State labels are given in cm$^{-1}$. Dashed lines are without repumping lasers, solid lines include $749$, $846$ and $855$-nm repumping lasers. Unknown Einstein $A_{21}$-coefficients were taken from the DREAM database \cite{Quinet.1999, Quinet.DREAM, Palmeri.2025}. (b) Experimental scattering rate without the infrared repumping lasers at a He pressure of $4.7\cdot 10^{-6}$\,Torr. Blocking the $450$-nm lasers for $2$ minutes pauses the optical population transfer. (c) Increase of the steady-state scattering rate with He-buffer-gas pressure in the $313$-nm cycle.
\label{fig:sim-450}}
\end{figure}
When searching for a closed optical cycle through the ground state at a convenient wavelength, transitions near $450$\,nm appear to be an attractive choice. Transitions at $448$ and $453$\,nm connect the even-parity $22\,308.82$-cm$^{-1}$ state ($4f^{12}5d6s$, $J=4$) to the odd-parity ground and $236.95$-cm$^{-1}$ states, respectively. According to the DREAM database, the $22\,308.82$-cm$^{-1}$ state decays into an odd-parity state at $8\,957.47$-cm$^{-1}$ through an $E1$ transition at $749$\,nm with an Einstein $A$ coefficient of $14.4\cdot 10^3$\,s$^{-1}$\,\cite{Quinet.1999, Quinet.DREAM, Palmeri.2025}. Because this state has the same parity as the ground states, $E1$ decays back to the cooling cycle are forbidden. Based on the electron $g$ factor and angular momentum geometry, we estimate a lifetime of $\approx\!90$\,ms for this state, which implies the need for repumping lasers.

In practice, however, we did not observe any evidence of population accumulating in this state. A possible explanation is that even though, initially, population is going through this decay channel, after a few optical cycles all population is leaking into $J\geq5$ states with lifetimes much longer than $90$\,ms. With an Einstein $A$ coefficient of $24.2\cdot 10^3$\,s$^{-1}$ at $4.8\,\upmu$m, the $E1$ transition into the $20\,228.75$-cm$^{-1}$~[$(\sfrac{7}{2},\sfrac{3}{2})^\mathrm{o}_5$] state is about twice as strong. The decay rate of this state back to the ground state is in the order of seconds\,\cite{Quinet.1999, Quinet.DREAM, Palmeri.2025}. Additionally, this state also decays in the order of seconds into an even longer lived $4f^{12}6s^2\,{}^3\mathrm{H}_6$ state at $12\,457.29$\,cm$^{-1}$, whose characterization is the main result of this work. All population excited into $J\geq4$ states eventually ends up in this state, which can be repumped with $846.31$, $1152.58$, or $1286.75$-nm light. These repumping wavelengths correspond to the only viable $E1$ transitions out of this state to the $24\,273.20$~[$(\sfrac{7}{2},\sfrac{5}{2})_5^{\mathrm{o}}$], $21\,133.68$~[$(\sfrac{7}{2},\sfrac{5}{2})_6^{\mathrm{o}}$] states, and the $20\,228.75$-cm$^{-1}$ state, respectively. Based on the observed HFS spectra, the $846$-nm and $1153$-nm transitions appear to be stronger repumping channels than the $1287$-nm transition, in accordance with the seconds lifetime of the $20\,228.75$-cm$^{-1}$ state specified in the DREAM database \cite{Quinet.1999, Quinet.DREAM, Palmeri.2025}.

While several arrangements of laser frequency tones are possible for this cycle, the most practical approach appears to be driving the two strong hyperfine lines satisfying $\Delta F = \Delta J$ for each transition. The main drawback of this cooling scheme is the relative weakness of the $448$ and $453$-nm transitions, with estimated partial lifetimes of $1.1$ and $1.8\,\upmu$s, respectively~\cite{Quinet.1999, Quinet.DREAM, Palmeri.2025}. While we can detect a signal from a large cloud of ions, the resulting low scattering rate might not provide sufficient laser cooling power for a thermal cloud or sufficient optical signal for single-ion imaging. Conversely, the narrower linewidth of $2\pi\times 0.23$\,MHz associated with this cycle offers a lower Doppler-cooling temperature limit, thus making this cycle suitable for second-stage cooling, similar to the multi-stage cooling in neutral Thulium~\cite{Provorchenko.2023}.

We find that adding helium buffer gas greatly increases the observed signal in both 313 and 450-nm cycles. This experimental observation is consistent with the cycles being open, as the buffer gas could be quenching undiscovered, long-lived leakage channels. A second possibility is that the buffer gas simply aids in trapping, leading to a higher ion density. The problem of leakage in the 450-nm cycle is further illustrated by a rate-equation simulation performed with the \texttt{qspec} Python package~\cite{qspec}. As is shown in Figure~\ref{fig:sim-450}(a), it predicts rapid population loss without repumping. Without the infrared repumping lasers, the population is first driven into the $8957.47$-cm$^{-1}$ and $20\,228.75$-cm$^{-1}$ states on a sub-millisecond timescale. Both states continuously decay back into the ground state, while about $38$\,\% of the $20\,228.75$-cm$^{-1}$ state branches into the long-lived $12457.29$-cm$^{-1}$ state on a timescale of $20$\,s. The $22308.82$-cm$^{-1}$ state, which is the only state contributing to the fluorescence signal, is not populated significantly after approximately $5$\,ms. However, a small amount of less than $1$\,\% remains until the $12457.29$-cm$^{-1}$ state is populated. These simulations qualitatively agree with the experimental findings in Figure~\ref{fig:sim-450}(b), showing the slowly decaying fluorescence signal from this remaining population when the pump lasers are on, but no population loss when the light is blocked. The $20\,228.75$-cm$^{-1}$ state being the bottleneck explains why no effect of the $749$-nm laser on the fluorescence signal is observed when only the $12457.29$-cm$^{-1}$ state is repumped. Consequently, the full optical cylcling scheme is expected to require $749$, $846$ and $855$-nm repumping lasers, reaching steady-state population after about $30$\,ms, with $23$\,\% of the population contributing to the fluorescence signal. Discrepancies between simulation and experiment in the population transfer timescale and observed scattering rate suggest that additional $M1$ decay channels, not listed in the DREAM database, might need to be considered. Figure~\ref{fig:sim-450}(c) shows the increased fluorescence signal when using He-buffer gas, here measured with the $313$-nm transition; hinting at quenching of the metastable states.

\subsection*{Laser cooling at 313\,nm}
\label{sec:laser cooling at 313 nm}
The $313$-nm line is the strongest transition we measured, with an Einstein coefficient of $A_{21}=2\pi\times 16.9$\,MHz~\cite{NIST}. It connects the ground state to an even-parity $J=5$ state with a $4f^{12}5d6s$ electronic configuration at $31\,926.82$\,cm$^{-1}$. Its relatively large linewidth, together with its short transition wavelength, offers a high Doppler cooling rate.

A significant challenge, however, is that the excited state can undergo $E1$ decay to several potentially metastable states, requiring multiple repumping lasers for efficient optical cycling. A significant portion of this work was dedicated to identifying which repumping transitions are most critical for optical cycling and Doppler cooling.

Via intermediate $J=4$ states, the excited state population can end up in the $237$-cm$^{-1}$ level, in which case $453$~nm light is required to repump the population. Similar to the $450$-nm cycle, all population tends to accumulate in the $12457.29$-cm$^{-1}$ state, but at a much higher rate. From there, the same laser tones can be used to transfer the population back into the ground state. However, this repumping process proved to be too slow to implement laser cooling. Instead, repumping lasers from the states that are connected directly to the excited state of the $313$-nm transition are required.

From measurements of the $854.84$, $926.51$, and $1306.58$-nm lines and searches for the $1056.04$ and $1190.09$-nm lines, we can deduce that the dominant decay channels are into the $21\,133.68$ [$(\sfrac{7}{2},\sfrac{5}{2})_6^{\mathrm{o}}$] and the $24\,273.20$-cm$^{-1}$ [$(\sfrac{7}{2},\sfrac{5}{2})_5^{\mathrm{o}}$] states. The decay into the $20\,228.75$-cm$^{-1}$ [$(\sfrac{7}{2},\sfrac{3}{2})_5^{\mathrm{o}}$] state through the $854.84$-nm transition was about an order of magnitude weaker, and was barely sufficient to enable taking a HFS spectrum.

Based on $E1$ selection rules, some population from the $313$-nm cooling cycle is expected to decay into the $22\,457.51$-cm$^{-1}$ [$(\sfrac{7}{2},\sfrac{3}{2})_4^{\mathrm{o}}$] and $23\,524.09$-cm$^{-1}$ [$(\sfrac{7}{2},\sfrac{5}{2})_4^{\mathrm{o}}$] levels. We applied light at $854.69$, $904.43$, $1056.04$, and $1190.09$\,nm to repump population through different excited states within the $313$-nm cycle (see Figure~\ref{fig:levels}), but observed no measurable change in fluorescence. This indicates that either the corresponding decay channels are weaker than the $854.84$-nm channel, or the lifetimes of these intermediate states are too short to cause significant population trapping.

Another potential metastable $(6,1)_5$ state suggested by the DREAM database\,\cite{Quinet.1999, Quinet.DREAM, Palmeri.2025} is located at $16\,567.47$\,cm$^{-1}$. Based on its energy and selection rules, this level should decay to the ground state via an $E$1 transition at approximately $604$\,nm. However, driving the $1\,297.74$-nm transition connecting this state to the $24\,273.20$-cm$^{-1}$ level produced no observable change in the fluorescence signal.

Finally, we tested whether $803$-nm light could drive measurable population from the ground state to the $12\,457.29$-cm$^{-1}$ level through a magnetic-quadrupole ($M2$) or electric-octupole ($E3$) transition, but no effect was observed.

\subsection*{Laser cooling on other transitions}
\label{sec:laser cooling on other transitions}
A $370$-nm transition connecting the ground state to a $J=4$ state at $27\,254.42$\,cm$^{-1}$ (not shown in Figure~\ref{fig:levels}) is another candidate for laser cooling. While this was the first transition we observed in the HCL, it is estimated to be $5$ times weaker than the $313$-nm transition. Furthermore, it shares most of the decay pathways to higher-$J$ states and introduces new decay channels toward $J\leq3$ metastable states. The $313$-nm line is therefore considered a more feasible option.

While this work has focused on transitions from the ground state to $J=4$ and $J=5$ manifolds, other configurations are possible. For example, a recent UV spectroscopy study~\cite{DenHartog.2024} identified several promising transitions originating from the $8\,957.47$-cm$^{-1}$ state, including transitions to $J=3$ states near $363$ and $331$\,nm with high branching factors of $0.995$ and $0.971$, respectively. Although lasers at these specific wavelengths were not available for the present work, these transitions represent a compelling avenue for future investigation.

\subsection*{Sympathetic cooling with \Baq{138}}
\label{sec:sympathetic cooling with 138Ba+}
Our ion trap can reliably trap and laser cool large clouds of \Baq{138} ions. By repeatedly switching the trap potential between high and low values, ions can be selectively removed to produce a laser-cooled few-ion crystal or a single ion. In principle, simultaneously co-trapping \Baq{} and \Tmq{} should allow for the sympathetic cooling of \Tmq{} ions during \Baq{} laser cooling. While we have demonstrated co-trapping of large \Baq{} and \Tmq{} clouds via trap-door loading, attempts to crystallize the two species simultaneously, so far, resulted in the loss of all \Tmq{} ions. We attribute this behavior to the lower mass of the \Baq{} ions, which experience a larger ponderomotive force. Consequently, they tend to push the \Tmq{} ions outwards during laser cooling rather than uniformly mix with or integrate the \Tmq{} ions into the ion crystal. Because of this spatial separation, we tend to lose \Tmq{} ions faster when toggling the trap potential. A potential solution is to utilize ions of a more similar mass, such as $^{171}$Yb$^+$ for sympathetic cooling; however, this capability is currently not available in our experimental setup. An alternative approach is to first trap and crystallize \Baq{} and then introduce \Tmq{} without lowering the trap potential, either by resonant photo-ionization of neutral \Tm{} or by capturing \Tmq{} through collisions with the trapped \Baq{} ions.

While co-trapping both species, we observed that the $453$-nm laser, required for \Tmq{}, populates the $5d\,^2$D$_{5/2}$ state in \Baq{} on a timescale of seconds, diminishing the \Baq{} fluorescence signal. By adding a repumping laser at a wavelength of $614$\,nm to drive the $5d\,^2$D$_{5/2}\rightarrow 6p\,^2$P$_{3/2}$ transition, the \Baq{} fluorescence signal can be reliably recovered.

\section*{Results}
\label{sec:results}

\subsection*{Hyperfine spectroscopy}
\label{sec:hyperfine spectroscopy}
\begin{figure*}[!t]
\centering
\includegraphics[width=\textwidth]{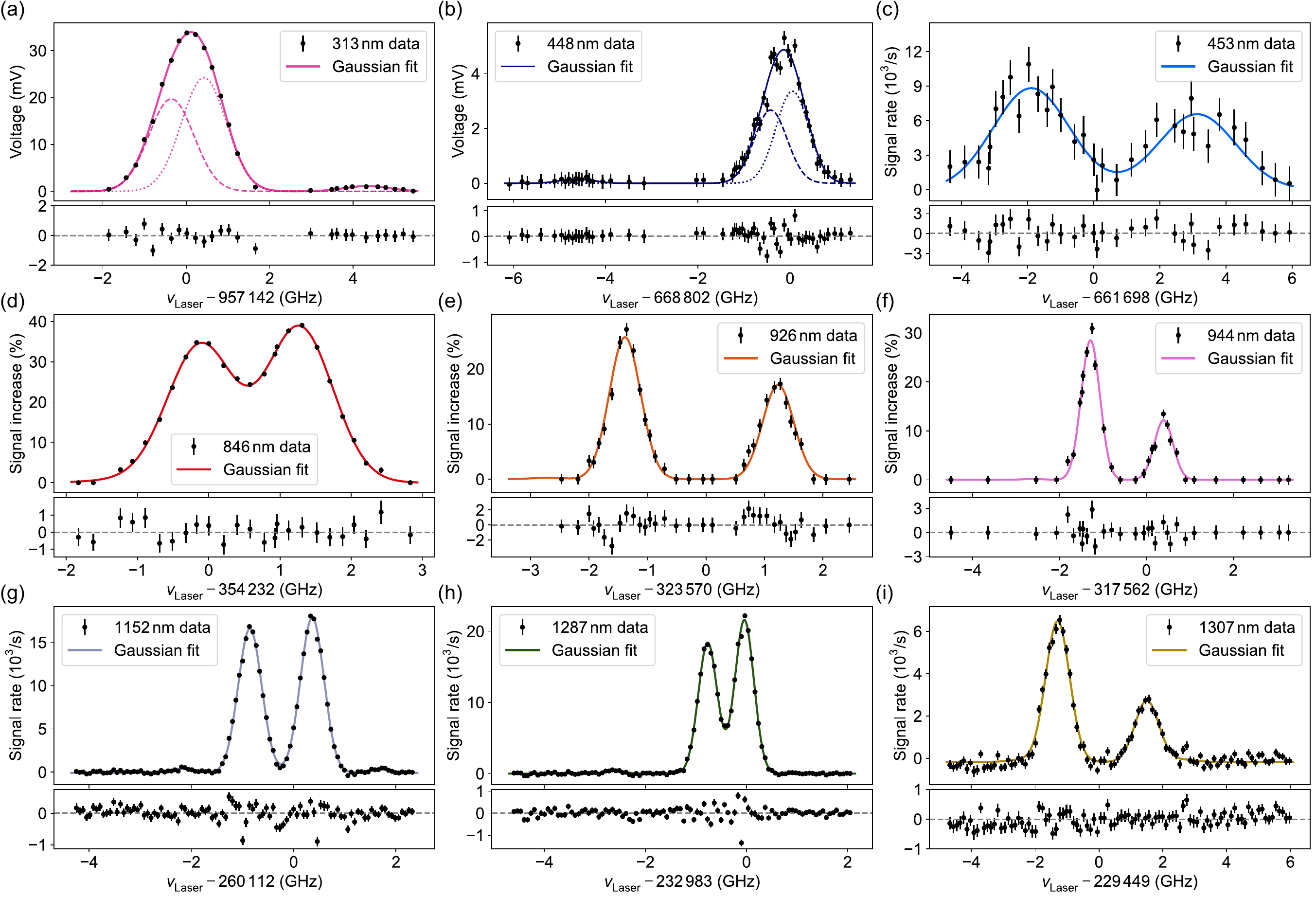}
\caption{\textbf{Hyperfine-structure (HFS) spectra of \Tmq{169}.} (a,b) were recorded using absorption spectroscopy with a hollow-cathode lamp, and (c--i) using fluorescence spectroscopy with an ionic cloud in a linear quadrupole trap. (c--f) A fresh ion cloud was produced for each data point; the laser light was generated using external-cavity diode lasers (ECDLs) and the laser frequencies were set to fixed values for each point. (g--i) Each spectrum was recorded with a single cloud of ions using the continuous-scan function of the optical parametric oscillator (OPO) laser.}
\label{fig:hyperfine-spectra}
\end{figure*}

A selection of fluorescence spectra that resulted from our measurement campaign is depicted in Figure~\ref{fig:hyperfine-spectra}.
From these spectra, the hyperfine-structure (HFS) $A$ constants of the excited states relevant to the proposed cooling schemes were determined. The extracted values are listed in Table~\ref{tab:hfs-constants}. Each spectrum was fitted with a sum-of-Gaussians model, using the HFS $A$ parameters of the lower and upper states as fit variables. Due to the large total electronic angular momentum, transitions where \mbox{$\Delta F \neq \Delta J$} are strongly suppressed. As a result, in several cases, only two spectral features were available to extract a single unknown $A$ constant, with the other fixed to a known reference value.
The specified uncertainties were determined from the least-square-fit uncertainties, adjusted for a reduced $\chi^2$ of one, as well as the statistics of repeated measurements and the frequency readout. A statistical $5$\,MHz frequency-readout uncertainty was added in quadrature (geometric sum) to the uncertainties of the determined resonance peak positions of spectra generated with the ECDLs, before calculating the weighted average over all measurements. The maximum between the standard errors of the unweighted and weighted average was then taken as the total statistical uncertainty. In cases where the $A$ constant was predominantly extracted from continuous scans of our OPO laser, a $2$\,MHz systematic frequency-calibration uncertainty was added linearly to the statistical uncertainty. This systematic uncertainty is the dominant contribution to our error budget. Other potential sources, such as light shifts, photon recoil shifts, and line-shape asymmetries, are negligible at the current level of precision. It should be noted that our uncertainty estimates are conservative, as performing metrological-grade high-precision measurements remains beyond the scope of this work.
\begin{table}[!t]
    \centering
    \caption{Hyperfine-structure (HFS) constants in \Tmq{169}. The doublet splittings are readily calculated using \mbox{$(J + \sfrac{1}{2})A$}. The HFS constants of the ground, $236.95$, and $31\,926.82$-cm$^{-1}$ states were kindly provided by Ref.\,\cite{Bodnar.2023}. The precise value of the $12\,457.29$-cm$^{-1}$ state was determined using microwave spectroscopy.}
    \begin{tabular}{cccc}
         \hline\hline
         Energy (cm$^{-1}$) & $J$ & $A$ (GHz) & Splitting (GHz) \\
         \hline
             0            & 4 & $-1.03913(2)$      & $4.67609(9)$    \\
           236.95        & 3 & $+0.29472(6)$      & $1.0315(2)$     \\
         12457.29        & 6 & $-0.3930(34)$      & $2.554(22)$     \\
                          &   & $-0.39230857(39)$  & $2.5500057(25)$ \\
         20228.75        & 5 & $-0.3311(29)$      & $1.821(16)$     \\
         21133.68        & 6 & $-0.2051(33)$      & $1.333(21)$     \\
         22308.82        & 4 & $-0.930(22)$       & $4.184(99)$     \\
         24273.20        & 5 & $-0.2110(31)$      & $1.160(17)$     \\
         31726.40        & 5 & $-0.5485(93)$      & $3.017(51)$     \\
         31926.82        & 5 & $-0.71871(5)$      & $3.9529(3)$     \\
         \hline\hline
    \end{tabular}
    \label{tab:hfs-constants}
\end{table}
The $A$ constants for the ground state and the $236.95$ and $31\,926.82$-cm$^{-1}$ states were obtained from collinear laser spectroscopy (CLS) measurements at Technische Universit\"at Darmstadt\,\cite{Bodnar.2023} and were used as starting values for determining the remaining constants. The HFS $A$ constants of the two low lying states from CLS as well as the $A$ constant of the $22308.82$-cm$^{-1}$ state, determined in this work, agree within $1\sigma$ with the values available in the literature \cite{Kebapci.2024}. All HFS splittings lie between $1$-$5$\,GHz, making \Tmq{} an ideal candidate for microwave-based qubit operations with well-separated HFS states.

\subsection*{12457.29-cm$^{-1}$ level as a qubit candidate}
\label{sec:12457.29-cm-1 level as a qubit candidate}
We observe experimentally that both of the optical cycles investigated lead to population leaking into the $12\,457.29$-cm$^{-1}$ state, which makes this state a natural choice for the metastable state in the $omg$ qubit protocol. From now on, we will refer to this state as the ``Gollum'' state. So far, the fastest way to transfer the ground-state population to this level is by applying $313$-nm light, which pumps the population into the state via its strong leakage channels. A population imbalance between the $F = \sfrac{11}{2}$ and $\sfrac{13}{2}$ sublevels can be created by coupling one of them to a different level, e.g., with $846$, $1153$, or $1287$-nm light. Addressing both HFS sublevels with these lasers will fully recover the population from this level.
\begin{figure}[!t]
\centering
\includegraphics[width=\linewidth]{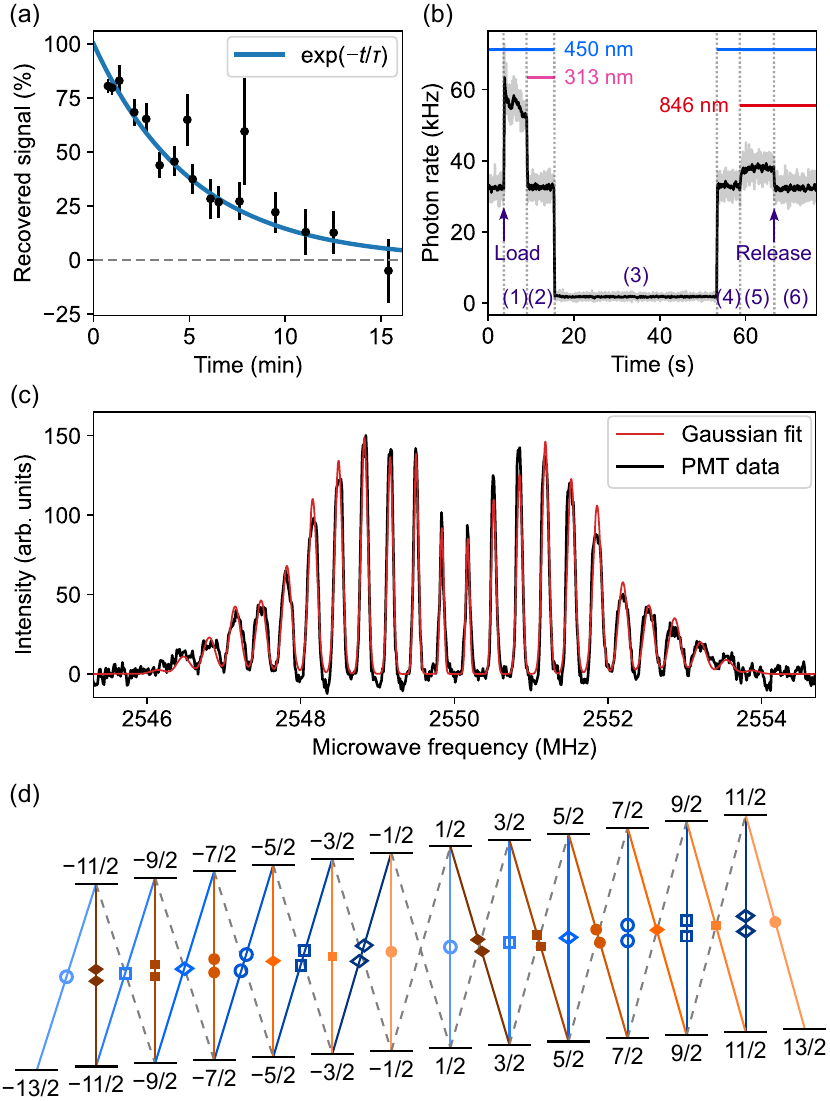}
\caption{\textbf{Microwave spectroscopy of the 12457.29-cm$^{-1}$ state.} The decay curve, pump-and-probe experiment, microwave spectrum and magnetic substructure of the $12457.29$-cm$^{-1}$ [$^3\mathrm{H}_6$] state (Gollum). (a) Population in this state as a function of time. An exponential function was fitted to the data, yielding a decay constant of $5.2(3)$\,min into the $^3\mathrm{F}_4$ ground state through $M2$ or $E3$ decay. (b) Example of a pump-and-probe measurement used to determine the lifetime of the $12457.29$-cm$^{-1}$ state. The numbers in parentheses refer to the sequence steps described in Methods. (c) Zeeman-resolved microwave spectrum of the transitions between the two hyperfine states of the $12457.29$-cm$^{-1}$ level, measured at a magnetic field of $1.34$\,G. A sum-of-Gaussians model was fitted to the data, with peak widths proportional to the peak's Zeeman shifts, to account for the variation of the magnetic field across the ion cloud. (d) Pairwise degenerate transitions, which contribute to the twelve inner peaks in the spectrum, are shown as solid lines, with the degenerate pair having the same color and labeled by the same geometric features, e.g., $\sfrac{13}{2}\rightarrow \sfrac{11}{2} $ is degenerate with $-\sfrac{1}{2}\rightarrow -\sfrac{1}{2}$. Non-degenerate transitions, which contribute to the twelve outer peaks, are shown as dashed gray lines. \label{fig:gollum-zeeman}}
\end{figure}
To test if the Gollum state lives long enough to be used as a qubit state, we measured its decay time using a pump-probe approach, with the result shown in Figure~\ref{fig:gollum-zeeman}(a). In this measurement, the $313$-nm laser is used to ``pump'' the population into the metastable state, while the closed $450$-nm cycle serves as the ``probe''. The $846$-nm laser is used as a repumper to return the population from the Gollum state back into the $450$-nm probe cycle for detection. The fractional population of the Gollum state at the time of the repumping process is equal to the fractional ``recovered signal'' shown on the $y$-axis of Figure~\ref{fig:gollum-zeeman}(a) and given by Eq.\,\eqref{eq:gollum-population} in Methods. This measurement scheme is shown in Figure~\ref{fig:gollum-zeeman}(b) and described in more detail in Methods.

Fitting an exponential decay model to the resulting decay curve yields a minimum lifetime of $5.2(3)$\,min. For the fit, it was assumed that immediately after turning off the $313$-nm laser, all of the population is in the Gollum state and for an infinite time after, all of the population has decayed back to the ground state. The determined lifetime is sufficiently long to prepare a qubit and execute quantum gates~\cite{DiVincenzo.2000, Blatt.2004}.

The high angular momentum of $J=6$ is a sufficient condition for the implementation of absorption-emission codes (\AE~codes), as proposed in Refs.\,\cite{Jain.2024, Aydin.2025}. In addition, the Gollum doublet contains an $F=\sfrac{13}{2}$ state, which is the minimum required for a the Clifford-friendly spin code described in Ref.\,\cite{Gross.2021}, which has been shown to be an absorption-emission code\,\cite{Aydin.2025,Jain.2024}.
These codes are robust against first-order absorption and emission errors but require a long state lifetime for operation. For the preparation of any qubit, the magnetic sublevels of the state need to be considered, which we analyzed with microwave spectroscopy. For this, the $1153$-nm laser was resonant with either of the $\sfrac{11}{2}\rightarrow \sfrac{11}{2},\sfrac{13}{2}$ transitions, such that only the $\sfrac{11}{2}$ state is repumped through the $21\,133.68$\,cm$^{-1}$ level. The fluorescence signal is taken from the light scattered in the $450$-nm cycle, such that increasing the population recovery rate from the Gollum state increases the signal. By applying resonant microwaves, the previously unaddressed population in the $F_+=\sfrac{13}{2}$ state is transferred into the $F_-=\sfrac{11}{2}$ state, where it can be repumped by the $1153$-nm laser. The resulting spectrum is shown in Figure~\ref{fig:gollum-zeeman}(c) with the hyperfine structure splitting between the $F_- = \sfrac{11}{2}$ and $F_+=\sfrac{13}{2}$ state of Gollum being $2.55$\,GHz and $\sfrac{11}{2}$ having a higher energy. The separable and strong microwave resonances show that the Gollum state is well-suited for microwave-based gates.

The observed microwave spectrum contains fewer than $3\times (2F_- +1)=36$ peaks that one would naively expect based on the linear Zeeman effect for two levels with positive Landé $g$-factors. This is due to the fact that \Tmq{169} has a nuclear spin of $I=\sfrac{1}{2}$: neglecting the contribution of the nuclear magnetic dipole, the total $g$-factors of the $F_{\pm}=J\pm \sfrac{1}{2}$ states are
\begin{align}
    g_+ &= g_J\dfrac{J}{J+1/2}\ \ \  \text{and}\ \ \ 
    g_- = g_J\dfrac{J+1}{J+1/2},
\end{align}
where $g_J$ is the electronic $g$-factor. The fact that $\sfrac{g_-}{g_+}=\sfrac{(J + 1)}{J}$ leads to peculiar degeneracies in the microwave transitions between the $m_F$ states of the two HFS levels: each $\pi$-transition \mbox{$m^\pi \rightarrow m^\pi$} is degenerate with a single $\sigma^\pm$ transition \mbox{$m^{\pm} \rightarrow m^{\pm}  \pm 1$} for which $m^\pi = m^{\pm}  \pm J$. On top of that, the degenerate transitions align by strength: the strongest $\pi$~transitions $m_F = \pm \sfrac{1}{2} \rightarrow m_F' = \pm \sfrac{1}{2}$ are degenerate with the strongest $\sigma$~transitions $m_F = \mp \sfrac{13}{2}\rightarrow m_F' = \mp \sfrac{11}{2}$; the second strongest $\pi$~transitions $m_F = \pm \sfrac{3}{2}\rightarrow m_F' = \pm \sfrac{3}{2}$ are degenerate with the second strongest $\sigma$~transitions $m_F = \mp\sfrac{11}{2}\rightarrow m_F'=\mp \sfrac{9}{2}$, etc, which is illustrated in Figure~\ref{fig:gollum-zeeman}(d). This results in $2\times(2F_-+1)=24$ distinct peaks, with the peaks corresponding to the weakest transitions barely visible. The lower observed heights of the central peaks corresponding to the strongest transitions still remain a puzzle.

Note that these degeneracies are not a unique feature of the Gollum state, but of all \Tmq{169} levels, or in fact any isotope with $I=\sfrac{1}{2}$ and an integer non-zero $J$. While the full impact of this feature on SPAM in \Tmq{169} and similar isotopes is yet to be determined, it can at least be utilized during state preparation to minimize the required number of frequency tones. Taking the Gollum state as an example and defining $\ket{0} \coloneqq \ket{\sfrac{13}{2}, -\sfrac{1}{2}}$ and $\ket{1} \coloneqq\ket{\sfrac{11}{2}, \sfrac{1}{2}}$, the qubit can be initialized in the $\ket{1}$ state with microwave-assisted optical pumping~\cite{tretiakov2024manipulating} through a combination of resonant laser and microwave tones. For example, addressing only the $F=\sfrac{13}{2}$ manifold inside a $313$-nm optical cycle with a single tone of $846$~nm light, while using unpolarized microwaves at the five tones that drive the $\pi$-transitions from $m=\sfrac{-9}{2}$ through $\sfrac{-1}{2}$, as well as the single tone that drives $\ket{\sfrac{11}{2},\sfrac{-11}{2}}\leftrightarrow \ket{\sfrac{13}{2},\sfrac{-9}{2}}$, will make $\ket{1}$ the only dark state in the cycle and force all population there.

\section*{Discussion}

\subsection*{Future directions}
\label{sec:future directions}
Despite driving nominally closed optical cycles, we have not observed any evidence for laser cooling of \Tmq{169}. We attribute this to ion shelving into intermediate odd-parity states for millisecond timescales, which severely limits the overall photon scattering rate. Mitigating this shelving, potentially by adding missing repumping lasers, could increase the cooling rate by a factor of up to $10^4$. This interpretation is supported by collateral measurements using a He-buffer gas, which quenches these metastable levels. To achieve laser cooling, we propose increasing the active cycling time by adding the necessary repumpers and improving the fluorescence-detection sensitivity to enable measurements on smaller ion numbers.

To increase the laser-cooling rate, first, the already known leakage channels of the $313$-nm cycling scheme can be closed, using two frequency tones at $846$, $854.8$, $927/944$ and $1307$\,nm simultaneously. Even if this turns out to be insufficient, it will facilitate the search for the remaining repumping transitions as the population will be diverted into the unknown weak decay channels. In the $450$-nm cycle, the $749$-nm transition is the only feasible unknown decay channel and it is expected to be strong enough to be relevant \cite{Quinet.1999, Quinet.DREAM, Palmeri.2025}. While He-buffer-gas quenching helped finding unknown repumping transitions, in the case of the $749$-nm transition, the estimated lifetime of its lower state ($8957.47$\,cm$^{-1}$) of $\approx\!90$\,ms might be reduced below the mean time of $\approx\,0.63$\,ms the upper state takes to decay into it. Consequently, adding the repumping laser in He-buffer gas would not improve the active optical cycling time, so that finding both hyperfine components simultaneously in a vacuum setup would be required.

\subsection*{Conclusions}
\label{sec:conclusions}
In conclusion, we have mapped out the hyperfine structure (HFS) required to assemble operational laser-cooling cycles in \Tmq{169}, combining absorption spectroscopy in a hollow-cathode lamp with fluorescence spectroscopy in an ion trap. The data include hyperfine $A$ constants for key excited states, validate a strong $313$-nm cycling transition, and identify a practical $448$/$453$-nm fluorescence-probing and optical-cycling scheme. These optical cycles are supported by the near-infrared repumping transitions identified in this work. Microwave spectroscopy in the long-lived $12457.29$-cm$^{-1}$ state, which we identified as a potential metastable qubit state, demonstrates the feasibility for microwave-based gates and state preparation.

Together, these results establish a closed HFS level scheme with MHz-level frequency references, providing the necessary data to constrain unwanted decays to metastable manifolds. This spectroscopic foundation reduces the required laser inventory and informs the design of compact, diode-based implementations. Future work will focus on quantifying branching ratios, demonstrating direct Doppler cooling, and leveraging the unique $I=\sfrac{1}{2}$, high-$J$ structure of \Tmq{169} for precision metrology and quantum-information primitives.

\section*{Methods}
\label{sec:methods}
The following section summarizes the basic theory and the experimental setups used for this study.

\subsection*{Theory}
\label{sec:theory}
To determine the hyperfine structure (HFS) constants and splittings of atomic states, the HFS shifts of individual levels must be parameterized. For $I=\sfrac{1}{2}$ the HFS shift $\Delta\nu_F$ of a level with quantum numbers $(J,F)$ relative to its center of gravity is \cite{Schwartz.1955}
\begin{align}
    \Delta\nu_F = \frac{A}{2}\!\left[F(F+1) - J(J+1) - I(I+1)\right],
\end{align}
so that, with the electric-quadrupole term $B$ vanishing, $A$ fully parameterizes the HFS. For an electric-dipole transition $J\!\to\!J'$, the allowed components satisfy $\Delta F=0,\pm1$ with $F=0 \nleftrightarrow F'=0$, yielding three lines when $J'=J\pm1$ and up to four when $J'=J$. The frequency of the $F\!\to\!F'$ component is
\begin{align}
    \nu_{F}^{F'} = \nu_{J}^{J'} + \Delta\nu^{F'} - \Delta\nu_F,
\end{align}
where $\nu_{J}^{J'}$ is the centroid transition frequency. Because the target manifolds are populated optically, the measured spectra also reflect population dynamics relevant to closing the cycles.
Here we determine the HFS $A$ constants of excited levels in \Tmq{169} that are not directly accessible from the ground state via optical dipole transitions; these levels are populated via resonant optical population transfer within the two candidate laser-cooling cycles at wavelengths of $450$ and $313$\,nm.

\subsection*{Absorption spectroscopy in a hollow-cathode lamp}
\label{sec:absorption spectroscopy in a hollow-cathode lamp}
Preliminary spectroscopy of \Tmq{} was performed in a hollow-cathode lamp (HCL) to locate strong optical transitions and measure their hyperfine structure. 
The HCL consisted of a thulium cathode mounted inside a quartz cell filled with neon buffer gas. A potential of $500$\,V was applied between the anode and the hollow cathode, creating a strong electric field that produces positively charged neon ions through electron-impact ionization. These ions are accelerated toward the cathode, sputtering \Tm{} atoms from its surface. Subsequently, \Tmq{} ions are produced through electron-impact ionization.

A laser beam was aligned through the HCL and terminated on a silicon amplified photodiode (Thorlabs PDA100A). To increase the optical density, the beam was passed through the HCL up to four times. When the laser frequency is tuned to a transition in \Tmq{} that exhibits a population imbalance between its lower and upper states, the transmitted power changes due to net stimulated emission or absorption. By scanning the laser frequency, a resonance peak or dip was recorded. To improve the signal-to-noise ratio, the laser was stepped across discrete frequency values while the HCL voltage was modulated using a high-voltage transistor driven by a rectangular waveform, effectively modulating the \Tmq{} density between zero and its maximum value. The same waveform was used for demodulation and averaging of the photodiode signal, analogous to a lock-in detection technique. A decrease in transmission was observed whenever the laser was on resonance and the \Tmq{} density was high. Using this method, we precisely determined the frequencies of the $313$, $346$, $370$, $448$, and $453$-nm transitions, which served as a starting point for subsequent ion-trap spectroscopy.

\subsection*{Fluorescence spectroscopy in an ablation cell}
\label{sec:fluorescence spectroscopy in ablation cell}
To evaluate the feasibility of producing \Tmq{} ions by laser ablation, we constructed a compact vacuum chamber for fluorescence detection. The chamber was pumped to a base pressure of $10^{-6}$\,mbar and consisted of a KF40 six-way cross equipped with UV-grade fused-silica viewports, connected to a vacuum pump and a helium buffer-gas inlet. A \Tm{} target, fabricated by pressing high-purity thulium metal into a cylindrical pellet ($1$\,cm in diameter), was mounted on one of the cross ports. A plume of \Tmq{} was generated by impinging the target with $1064$-nm laser pulses introduced through the opposite port.

The probe laser beam was aligned orthogonally to the ablation plume at a distance of approximately $1$\,cm from the target surface. Laser-induced fluorescence was collected by a lens placed inside the chamber, positioned one focal length of $2$\,cm from the intersection of the plume and the probe beam. The fluorescence photons were detected using a photomultiplier tube (PMT) and time-binned with a temporal resolution of down to $5$\,ns.

Although fluorescence was detectable even under vacuum conditions, the method performed best with buffer-gas pressures between $10^{-3}$ and $10^{-1}$\,mbar, using either helium or air. Measurements on the $313$-nm transition confirmed that laser ablation can reliably generate a high density of thulium ions, demonstrating the suitability of this approach for ion loading in subsequent trap experiments.

\subsection*{Fluorescence spectroscopy in a linear quadrupole trap}
\label{sec:fluorescence spectroscopy in a linear quadrupole trap}
\begin{figure}[!t]
\centering
\includegraphics[width=\linewidth]{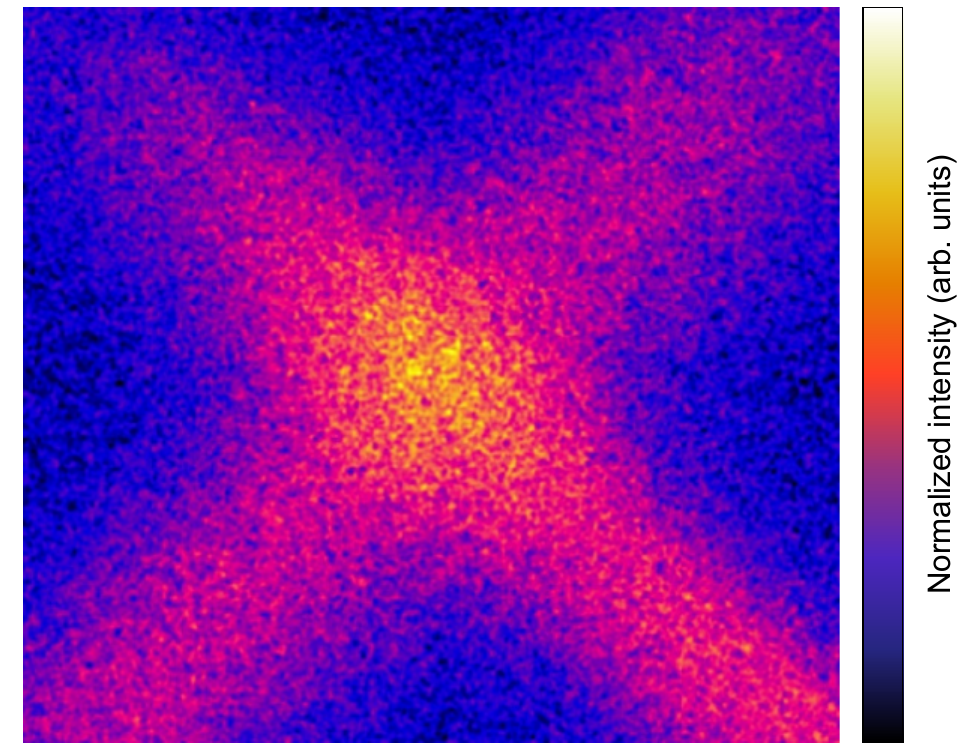}
\caption{\textbf{Image of an illuminated cloud of \Tmq{} ions.} The image was taken with an EMCCD camera. Both the $448$- and $453$-nm lasers cross the trap center twice, forming an X shape. The image was post-processed with a median and a Gaussian filter to reduce image noise and improve contrast.}
\label{fig:camera-tm}
\end{figure}
Most of our spectroscopy measurements were performed in a linear quadrupole trap originally designed for experiments with trapped \Baq{} ions~\cite{Putnam.2024, Putnam.2024.thesis}. The trap consists of four parallel, segmented rods arranged symmetrically around the trap axis at a distance of $1$\,cm. Each rod is divided into three electrodes. A radio-frequency (RF) voltage of $0.5$--$1$\,kV at $0.939$\,MHz is applied to a pair of opposing center segments to provide radial confinement. The fringe field of the RF potential also contributes to weak axial confinement; additional static voltages can be applied to the outer segments to displace the ion cloud or increase the axial trapping potential. The trap is situated inside a sealed chamber and can operate under ultrahigh vacuum conditions or with buffer gas.

\Tmq{169} ions are produced by ablating a pure thulium metal target with $532$-nm laser pulses, following a procedure similar to that used in the ablation cell. The RF amplitude is temporarily lowered $20$--$150$\,ms after the ablation pulse is triggered to capture ions from the ablation plume passing through the trap center.

The trapped ions are illuminated by multiple laser beams intersecting at the trap center, with up to two passes per beam. Each beam may contain several frequency components. The main cooling configuration for the $450$-nm colling cycle comprises two tones at $448$\,nm and two at $453$\,nm, crossing the trap center twice in an X-shaped geometry, as illustrated in Figure~\ref{fig:camera-tm}. For the $313$-nm system, two counter-propagating beams each intersect the trap center twice in a “plus” geometry; the second beam is frequency-shifted by $+722$\,MHz relative to the first using acousto-optic modulators (AOMs). The AOM shift ensures that both HFS levels are addressed. One infrared (IR) beam is counter-propagating the $450$-nm beam, and two additional IR beams intersect the trap center once.

Laser-induced fluorescence is collected by two four-lens imaging systems located above and below the chamber, projecting the image onto an EMCCD (Andor iXon3) and a CMOS camera (Thorlabs CS2100M), respectively. In addition, each optical path has an option to focus the light onto a photomultiplier tube (PMT) instead of the camera. For transitions involving energetically high-lying states, population trapping in long-lived metastable levels can substantially reduce the steady-state fluorescence. To counteract this effect, helium buffer gas at pressures between $10^{-5}$ and $10^{-3}$\,mbar is introduced to quench the population back to the ground state through collisional relaxation.

All fluorescence spectra were recorded by collecting photons from the main cooling transitions at $448$, $453$, or $313$\,nm. Potential repumping transitions were identified by monitoring their effect on this steady-state fluorescence. This technique works because the cooling lasers continuously populate intermediate ``dark'' states. When an additional laser is scanned into resonance with a repumping transition, it depopulates one of these metastable states and returns ions to the main cooling cycle. This results in a measurable increase in the steady-state fluorescence from the cooling transitions similar to the 846-nm signal increase observed in Figure~\ref{fig:gollum-zeeman}(b). A quantitative investigation of the transition strengths is challenging, as our detection scheme depends strongly on the internal population dynamics; therefore, such an analysis remains beyond the scope of this work. Nevertheless, we note that the ratios of the Einstein $A$ coefficients sourced from the DREAM database\,\cite{Quinet.1999, Quinet.DREAM, Palmeri.2025} are in good agreement with our experimental observations, e.g., in the case of the three infrared repumping transitions from the Gollum state.

\subsection*{Lifetime measurement of the 12457.29-cm$^{-1}$ level}
\label{sec:lifetime measurement of the 12457.29-cm-1 level}
The lifetime of the $12457.29$-cm$^{-1}$ level was measured with the experimental sequence shown in Figure~\ref{fig:gollum-zeeman}(b). The enumerated sequence steps are described in the following. (1): The sequence starts when \Tmq{169} is loaded through laser ablation.  (2): After a few seconds, the $313$-nm laser pumps ions out of the 450-nm cycle into the Gollum state, which decreases the detected photon rate down to the background level. (3): All lasers are turned off for a varying dark time, during which the population spontaneously decays into the ground state. The duration of this step corresponds to the $x$-axis values in Figure~\ref{fig:gollum-zeeman}(a). (4): The $450$-nm lasers are turned back on to measure the fraction of the population that decayed back to the ground state. (5): $846$-nm light recovers the population that remained in Gollum to the $450$-nm cycle. (6): The ions are released from the trap to determine the residual laser background.

The change in fluorescence upon adding $846$-nm light, normalized to the maximum fluorescence from all ions, yields the fractional population in the Gollum state given by
\begin{align}
p_\mathrm{Gollum} = 1 - \frac{1}{1 - \gamma}\frac{\Gamma_{(4)} - \Gamma_{(6)}}{\Gamma_{(5)} - \Gamma_{(6)}},
\label{eq:gollum-population}
\end{align}
where $\Gamma_{(i)}$ is the fluorescence rate in the sequence interval ($i$) and $\gamma$ is a correction, as the same number of ions produced about $\gamma \coloneqq 10$\,\% more fluorescence in the $450$-nm cycle when the $846$-nm laser was added, slightly reducing the resulting lifetime.

\subsection*{Microwave spectroscopy in a linear quadrupole trap}
\label{sec:microwave spectroscopy in a linear quadrupole trap}
By exchanging the spectroscopy laser with a microwave source in the single-digit GHz regime and applying a magnetic field to the trap, transitions between the Zeeman components of the two HFS states within a fine-structure manifold can be probed. The microwave spectrum in Figure~\ref{fig:gollum-zeeman}(c) was produced with a horn antenna. We were also able to record microwave spectra by applying the radio-frequency voltage to the trap electrodes directly. However, this method produced an asymmetric Zeeman spectrum, hinting at an imbalance between $\sigma^{+}$ and $\sigma^{-}$ microwave-polarization components.

\subsection*{Laser systems}
\label{sec:laser systems}
The wide range of wavelengths employed in this work requires different types of lasers. For infrared transitions from $1056$ to $1307$\,nm, we used an optical parametric oscillator (OPO, H\"{u}bner C-wave GTR). For transitions between $749$ and $944$\,nm, we used in-house built external-cavity diode lasers (ECDLs) or a Ti:sapphire laser (M-squared SolsTiS). Additional ECDLs were used to generate $448$ and $453$-nm light. The $313$-nm light was generated with a frequency-quadrupled diode laser (Toptica TA-FHG).
All laser frequencies were measured with a wavelength meter (Angstrom/HighFinesse WS-8). For fluorescence spectroscopy, we used laser powers of $0.05 - 10$\,mW with beam radii of $0.3 - 3$\,mm, and various polarizations.

\bibliography{bibliography}

\section*{Data availability}
The datasets generated and analyzed during the current study are available from the corresponding author on reasonable request.

\section*{Acknowledgements}
The authors acknowledge H. Bodnar, B. Cheal,  K. K\"onig and W. N\"ortersh\"auser for sharing unpublished hyperfine spectra and constants, as well as P. Palmeri for sharing unpublished data from the DREAM database.

\section*{Research funding}
This work was supported by National Science Foundation grants PHY-2207985 and OMA-2016245, NIST award number 60NANB24D104, and the Gordon and Betty Moore Foundation DOI: 10.37807/GBMF11566.

\section*{Author contributions statement}
P.M. and A.T. wrote the manuscript, prepared the figures, conducted the measurements, and analyzed the results. 
P.M. built the 448 and 453~nm ECDLs and performed calculations for Figure~2(a).
A.Y. and N.H. built the infrared ECDLs and the buffer gas supply system.
W.C.C. and P.H. conceived and supervised the experiment.
All authors reviewed the manuscript.

\section*{Additional information}
\textbf{Competing interests} The authors declare no competing interests.

\end{document}